\documentclass[twocolumn,showpacs,preprintnumbers,superscriptaddress,pra]{revtex4}
\usepackage{amsmath}
\usepackage{dcolumn}
\usepackage{bm}
\usepackage{epsfig}
\usepackage{graphicx,float}
\usepackage{amsfonts}
\usepackage{amssymb}
\usepackage{mathrsfs}
\usepackage{color}
\usepackage{appendix}

\begin{document}

\title{Phase-controlled Fano resonance by the nanoscale optomechanics}
\author{Jian-Qi Zhang}
\affiliation{State Key Laboratory of Magnetic Resonance and Atomic and Molecular Physics,
Wuhan Institute of Physics and Mathematics, Chinese Academy of Sciences - Wuhan National
Laboratory for Optoelectronics, Wuhan 430071, China }
\author{Yi Xu}
\affiliation{School of Physics and Electric Engineering, Guangzhou University, Guangzhou
510006, China}
\author{Keyu Xia}
\affiliation{ARC Centre for Engineered Quantum Systems, Department of Physics and
Astronomy, Macquarie University, NSW 2109, Australia}
\author{Zhi-Ping Dai}
\affiliation{Department of Physics and Electronic Information Science, Hengyang Normal
University, Hengyang 421008, China}
\author{Wen Yang}
\email[Corresponding author Email: ]{yangw@csrc.ac.cn}
\affiliation{Beijing Computational Science Research Center, Beijing 100084, China }
\author{Shangqing Gong}
\email[Corresponding author Email: ]{sqgong@ecust.edu.cn}
\affiliation{Department of Physics, East China University of Science and Technology,
Shanghai 200237, China}
\author{Mang Feng}
\email[Corresponding author Email: ]{mangfeng@wipm.ac.cn}
\affiliation{State Key Laboratory of Magnetic Resonance and Atomic and Molecular Physics,
Wuhan Institute of Physics and Mathematics, Chinese Academy of Sciences - Wuhan National
Laboratory for Optoelectronics, Wuhan 430071, China }

\begin{abstract}
Observation of the Fano line shapes is essential to understand
properties of the Fano resonance in different physical systems. We explore a tunable Fano resonance by
tuning the phase shift in a Mach-Zehnder interferometer (MZI) based on a
single-mode nano-optomechanical cavity. The Fano resonance is
resulted from the optomechanically induced transparency
caused by a nano-mechanical resonator and can be tuned by applying an optomechanical MZI. By tuning the phase
shift in one arm of the MZI, we can observe the periodically varying line shapes of the
Fano resonance, which represents an elaborate manipulation of the
Fano resonance in the nanoscale optomechanics.
\end{abstract}

\date{\today }
\maketitle


Fano profiles are typically asymmetric line shapes, resulted from
quantum interference between discrete energy states and a continuum
spectrum \cite{rmp-82-2257}. Since the celebrated discovery by Fano
\cite{Cimento-12-154}, the Fano profiles have been observed in
various physical systems with atoms \cite{PR-124-1866,
prl-107-163604, Nature-440-315}, photons \cite{pre-71-036626,
pre-74-046603, pra-79-013809, NL-9-1663}, and solid-state systems
\cite{PhysScr-74-259,rmp-79-1217, prl-90-084101}. Recently, the Fano
profile has also been studied based on the optomechanics
\cite{arxiv}, where a nano-mechanical resonator (NAMR) inside an
optical cavity interacts with the cavity mode via the radiation
pressure.

The field of the optomechanics has become a rapidly developing area
of physics and nanotechnology over the past decades. Since the NAMR
is very sensitive to the tiny external force, the study of the
optomechanics is mainly motivated by the precision measurement, such
as the ultra-sensitive detections of the mass \cite {ajp-95-2682}, the
charge number \cite{pra-86-053806}, the gravitational wave
\cite{prl-95-221101} and the displacement (or the force) \cite
{Physics-2-40,prl-108-120801}. Moreover, there have been extensive
interests in observing quantum properties of the NAMRs, including
optomechanical entanglement
\cite{prl-88-120401,prl-98-030405,prl-101-200503}, photon blockade
\cite{prl-107-063601,prl-107-063602}, carrier-envelope
phase-dependent effect \cite{OL-38-353}, and optomechanically
induced transparency (OMIT) \cite{pra-81-041803,Science-330-1520,
pra-86-013815}. By adding more continuum spectra into the Fano resonance system,
one can gain more degrees of freedom to manipulate the resonance interaction\cite{pra-84-033828}.

In this Letter, we focus on a tunable Fano profile generated by
the quantum interference between a quasi-continuum light and an output
light from the optomechanics, via the NAMR-induced OMIT. Such quantum interference
is fulfilled by the MZI constructed by an optomechanical system and a phase shifter.
The key point of our scheme is the phase tunable Fano profile, which is based on the unique characteristic of the OMIT.
The output light from the optomechanics involves both symmetric and
asymmetric components. Due to this reason, the Fano profile
can be adjusted by tuning the phase in the second continuum light rather than
by tuning the input light frequency\cite{arxiv, oe-21-6601}.
Moreover, our scheme presents an ideal case to control the Fano resonance of the optomechanical
system by using a phase shifter outside the optomechanical system.
Furthermore, compared with the
conventional Fano profiles exhibiting sharp asymmetric Fano profiles
in transmission or absorption \cite {rmp-82-2257}, the Fano profile
in our scheme can be tuned periodically, i.e., from a symmetric line
shape \cite{pra-84-033828} to an asymmetric one, and then back to a
symmetric one. Inverted Fano profiles can be obtained by properly manipulating the phase
shifter. We emphasize that, since there are some additional quantum interferences in optomechanics for the OMIT effect,
our scheme based on the Mach-Zehnder interferometer (MZI) involves
more interferences than in the conventional MZI, which might be
practical for the precision measurement using NAMR-based
optomechanics.

\begin{figure}[tbp]
\includegraphics[width=6cm]{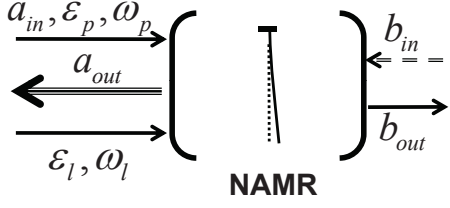}\newline
\caption{Schematic diagram of an optomechanical system involving a
NAMR. The cavity mirrors are highly reflected so that the cavity decay
is small. $b_{in}$ represents the input of a vacuum field, $a_{out}$
and $b_{out}$ indicate the reflected and transmitted lights from the
optical cavity. Other denotations are defined in the text.}
\label{inoutput}
\end{figure}

The core part in our scheme is an optomechanical system sketched in
FIG.\ref {inoutput}, where a NAMR is suspended in an optical cavity
composed of two fixed mirrors with identical finite transmission.
The cavity mode, which is driven by a strong external field with the
coupling strength $\varepsilon_{l}$ and the frequency $\omega_{l}$,
interacts with the NAMR by a radiation pressure coupling. In
addition, a weak probe light with the frequency $\omega_{p}$ is
incident into the optical cavity, which is of a large bandwidth and
can be considered as a quasi-continuum light with the OMIT effect.
Then, in the rotating frame with the driving field frequency
$\omega_{l}$, the Hamiltonian for this system reads
\begin{equation}
\begin{array}{lll}
\hat{H} & = & -\hbar \Delta _{c}c^{\dag }c+(\dfrac{p^{2}}{2m}+\dfrac{m\omega
_{m}^{2}}{2}q^{2}) \\
& - & \hbar gqc^{\dag }c+(i\hbar \varepsilon _{l}c^{\dag
}+\text{\textrm{ H.c. }}),
\end{array}
\label{W1}
\end{equation}
where $\Delta _{c}=\omega _{c}-\omega _{l}$ is the detuning of the
driving field frequency $\omega_{l}$ from the bare cavity frequency
$\omega_{c}$, $c$ is the bosonic annihilation operator of the cavity
mode, $p$ and $q$ are the position and momentum operators of a
NAMR with the mass $m$ and the frequency $\omega_{m}$,
$g=\omega_{c}/L$ is the radiation pressure coupling between the
cavity field and the NAMR with $L$ being the
cavity length \cite{pra-85-021801}.

A proper analysis of the system should involve the dissipations, such
as the photon loss from the cavity and the Brownian noise from the
environment. So the dynamics of the system is governed by the
nonlinear quantum Langevin equations
\begin{equation}
\begin{array}{lll}
\frac{\partial q}{\partial t} & = & p/m, \\
\frac{\partial p}{\partial t} & = & -m\omega _{m}^{2}q+\hbar gc^{\dag
}c-\gamma _{m}p+\sqrt{2\gamma _{m}}\xi (t), \\
\frac{\partial c}{\partial t} & = & -[2\kappa +i(\Delta
_{c}-gq)]c+\varepsilon +\sqrt{2\kappa }(a_{in}+b_{in})\,,%
\end{array}
\label{W2}
\end{equation}
where $\gamma _{m}$ and $\kappa $ are introduced as the
NAMR decay and the cavity decay, respectively.
The quantum Brownian noise $\xi $ comes from the coupling between
the NAMR and the environment. The correlation
of the noise operator is $\langle\xi (t)\xi (t^{\prime})\rangle
=N_{th}\delta (t-t^{\prime })$ with
$N_{th}=1/(e^{\frac{\hbar\omega_{m}}{k_{B}T}}-1)$ at a temperature
$T>\hbar \omega_{m}/k_{B}$. Eq.(\ref{W2}) can be solved after all
operators are linearized as the steady-state mean values plus the
small fluctuations,
\begin{equation}
q=q_{s}+\delta q,~~~p=p_{s}+\delta p,~~~c=c_{s}+\delta c\,,  \label{W3}
\end{equation}
with $\delta q$, $\delta p$ and $\delta c$ being the small
fluctuations around the corresponding steady-state values $q_{s}$,
$p_{s}$ and $c_{s}$, respectively.

After substituting Eq.(\ref{W3}) into Eq.(\ref{W2}), ignoring the second-order
small terms, and introducing the Fourier transforms $f(t)=1/2\pi
\int_{-\infty }^{+\infty }f(\omega)e^{-i\omega t}d\omega$, we obtain the
steady values $p_{s}=0$, $q_{s}=\hbar g|c_{s}|^{2}/(m\omega
_{m}^{2})$, and $c_{s}=\varepsilon _{l}/(2\kappa +i\Delta )$ with
$\Delta =\Delta_{c}-gq_{s}$. Then the solution of $\delta c$ is
identical to the one in Ref. [25],
\begin{equation}
\begin{array}{lll}
\delta c(\omega ) & = & V(\omega )\xi (\omega )+E(\omega )[a_{in}(\omega
)+b_{in}(\omega )] \\
& + & F(\omega )[a_{in}^{\dag }(-\omega )+b_{in}^{\dag }(-\omega )],
\end{array}
\label{W6}
\end{equation}
with
\begin{equation}  \label{W7}
\begin{array}{lll}
V(\omega ) & = & igc_{s}[2\kappa -i(\Delta +\omega )]/d(\omega ), \\
E(\omega ) & = & \sqrt{2\kappa }\{m[2\kappa -i(\Delta +\omega )](\omega
^{2}-\omega _{m}^{2}+i\omega \gamma _{m}) \\
& - & i\hbar g^{2}|c_{s}|^{2}\}/d(\omega ), \\
F(\omega ) & = & -i2\sqrt{2\kappa }\hbar g^{2}c_{s}^{2}/d(\omega ), \\
d(\omega ) & = & m[\Delta ^{2}+(2\kappa -i\omega )^{2}](\omega ^{2}-\omega
_{m}^{2}+i\omega \gamma _{m}) \\
& + & 2\hbar g^{2}|c_{s}|^{2}\Delta.
\end{array}
\end{equation}

According to the input-output relation of the cavity
\cite{D.F.Walls-Quantum Optics}, the output field from the port
$b_{out}$ is given by \cite {Science-330-1520}
\begin{equation}
\begin{array}{lll}
b_{out}(\omega ) & = & b_{in}(\omega)-\sqrt{2\kappa}\delta c(\omega)
\\ & \simeq & -\sqrt{2\kappa}E(\omega)a_{in}(\omega),
\end{array}
\label{W8}
\end{equation}
where we have assumed that the input noises regarding $b_{in}$,
$a_{in}^{\dag}$ and $\xi$ are negligible compared to the dominant
contribution from the input channel $a_{in}$ to the cavity \cite
{pra-81-041803,pra-85-021801,pra-86-053806}.

\begin{figure}[tbph]
\includegraphics[width=8cm]{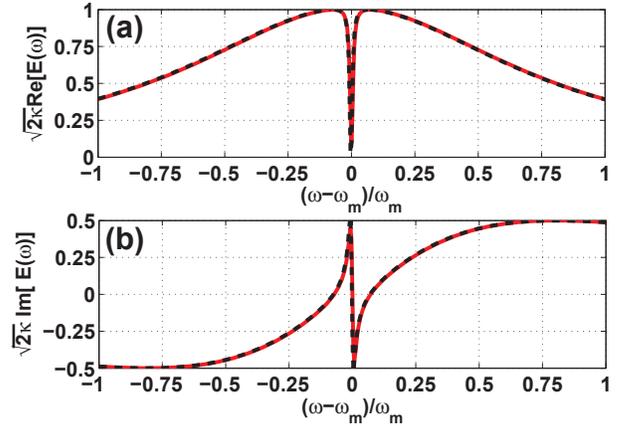}
\caption{(Color online) (a) The real component $\sqrt{2\protect\kappa}
Re[E(\protect\omega)](=\protect\sqrt{2\protect\kappa}\mathbf{\Re}[
E(\protect\omega)])$ of the transmission versus the normalized
detuning $(\protect\omega - \protect\omega_m)/\protect\omega_{m}$.
(b) The imaginary component $\protect\sqrt{2\protect\kappa}Im[
E(\protect\omega)](=\protect\sqrt{2\protect\kappa}\mathbf{\Im}[
E(\protect\omega)])$ of the transmission versus the normalized
detuning $(\protect\omega -
\protect\omega_{m})/\protect\omega_{m}$. The red solid lines (black
dotted lines) represent the exact (approximate) solution using
Eq.(\ref{W7}) [Eq.(\ref{W81})]. We choose $\protect\lambda _{c}\equiv 2
\protect\pi c/\protect\omega _{c}=400$ nm, $L=1$ mm, $m=1.45$ ng,
$\protect\kappa =2\protect\pi \times 10$ MHz, $\protect\omega _{m}=2
\protect\pi\times 30$ MHz, $\protect\gamma _{m}=2\protect\pi \times
3$ kHz, $ \Delta = 0.998\protect\omega_m$ and $\protect\varepsilon
_{l}=\protect\sqrt{ 2P\protect\kappa /\hbar\protect\omega_{c}}$ with
$P=5$ mW \protect\cite {pra-81-041803, Nature-460-724,
Nature-452-72}.} \label{symasym}
\end{figure}

To further understand the transmission $b_{out}$, we employ the
following conditions \cite{pra-81-041803, Science-330-1520}: (i)
$\Delta\simeq\omega_{m}$ and (ii) $\omega _{m}\gg\kappa$. The first
condition means that the optical cavity is driven by a red-detuned
laser field which is on resonance with the optomechanical
anti-Stokes sideband. The second condition is the well-known
resolved sideband condition, which ensures a distinguishable
splitting of the normal mode \cite{Science-330-1520}. Moreover, we
assume $\omega\simeq\omega _{m}$ so that there is a strongest
coupling between the NAMR and the cavity \cite{pra-81-041803} with
$\omega^{2}-\omega_{m}^{2}\simeq 2\omega_{m}(\omega -\omega _{m})$.
As a result, $E(\omega)$ in Eq.(\ref{W7}) can be reduced to
\begin{equation}
E(\omega )\simeq\dfrac{\sqrt{2\kappa }}{2\kappa -i(\omega -\omega
_{m})+\beta /[\gamma_{m}/2-i(\omega -\omega _{m})]}\,,  \label{W81}
\end{equation}
with $\beta =\dfrac{\hbar g^{2}|c_{s}|^{2}}{2m\omega_{m}}.$

When the driving field and the quasi-continuum probe light are
simultaneously incident into the optomechanical cavity, only the
probe light with the frequency at $\omega_{p}=\omega_{c}+\omega
_{m}$ can be reflected from the output port $a_{out}$, and the rest
transmits to the port $b_{out}$. The corresponding transmission
$\sqrt{2\protect\kappa} E(\protect\omega )=-b_{out}(\omega)/a_{in}
(\omega)$ involves a symmetric real part (with the even symmetry)
and an asymmetric imaginary part (with the odd symmetry) (see
FIG.\ref{symasym}). The good agreement between the results by FIG.\ref{W7}
and Eq.(\ref{W81}) indicates that the transmitted light is resulted from
the OMIT effect \cite {pra-81-041803, Science-330-1520}. This output
light from the optomechanics can be treated as a quasi-continuum
light excluding a single-mode component at the frequency
$\omega_{l}-\omega_{m}$. To the best of our knowledge, this kind of
output light for the phase-tuning Fano profile is an unique
characteristic of our model involving the optomechanics and has
never been reported previously in other physical systems.

\begin{figure}[tbp]
\includegraphics[width=8cm]{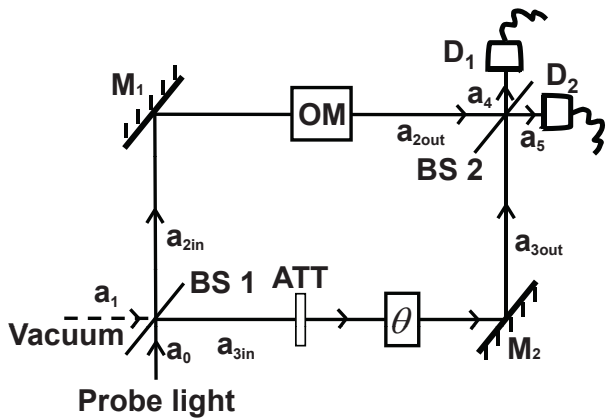}
\caption{Mach-Zehnder interferometer (MZI) involving an optomechanical
cavity (OM). In the upper arm, the light is reflected by the mirror
M$_{1}$ and then passes through the OM. The light in the lower arm
first passes through an attenuator (ATT), and then experiences a
phase shift $\protect\theta$. The two light beams interfere in the
second beam splitter (BS2) at the output port before they are detected by the
detectors D$_{1}$ and D$_{2}$.} \label{Mz}
\end{figure}

We present below how to control the Fano profile via phase-tuning
in one of the arms of the MZI. The setup for such tunable Fano profiles
is presented in FIG.\ref{Mz}, where the probe light is
a quasi-continuum field. After this probe light $a_{0}(\omega)$
along with the vacuum field $a_{1}(\omega)$ is sent into the first
beam splitter (BS1) of the MZI, the two fields are split into two
superposition fields as \cite{quantumoptics},
\begin{equation}
\begin{array}{c}
a_{2in}(\omega )=[a_{0}(\omega )+ia_{1}(\omega )]/\sqrt{2}, \\
a_{3in}(\omega )=[a_{1}(\omega )+ia_{0}(\omega )]/\sqrt{2}\,.
\end{array}
\label{W9}
\end{equation}

We assume that the optical-path difference between the two arms of the
MZI is integral times of the wavelength. When the light beam in
$a_{2in}(\omega)$ [=$a_{in}(\omega)$ in FIG.\ref{inoutput}] passes
through the optomechanical system, the output light from the
optomechanics owns both symmetric real and asymmetric imaginary
components. On the other hand, the quasi-continuum light in the
lower arm $a_{3in}(\omega)$ passes through an attenuator (ATT) and
experiences a phase shift $\theta$. Thus the modes $a_{2in}(\omega
)$ and $a_{3in}(\omega)$ can be written as
\begin{equation}
\begin{array}{cll}
a_{2out}(\omega ) & = & -\sqrt{2\kappa }\mu E(\omega )[a_{0}(\omega
)+ia_{1}(\omega )]/\sqrt{2}, \\
a_{3out}(\omega ) & = & e^{i\theta }[a_{1}(\omega )+ia_{0}(\omega
)]/\sqrt{2},
\end{array}
\label{W10}
\end{equation}
where $\mu^{2}=|a_{2out}(\omega)/a_{3out}(\omega)|^{2}$ depends on
the amplitude transmission of the optomechanics as well as the ATT
absorption in the MZI. The output light $a_{2out}(\omega)$ is much
weakened due to the highly reflected mirror of the optomechanical
cavity, and excludes a single-mode component at the frequency
$\omega_{l}-\omega_{m}$. The other light $a_{3out}(\omega )$ is a
quasi-continuum light, which is also weak after experiencing the
ATT. If we assume the light $a_{2out}(\omega)$ to be of the unit
intensity, the phase-shifted light $a_{3out}(\omega)$ is of the
intensity $1/\mu^{2}$. The two lights interfere in the second beam
splitter (BS2) before they are measured by the detectors. The corresponding
output lights are given by
\begin{equation}
\begin{array}{cll}
a_{4}(\omega ) & = & -\sqrt{2\kappa }\mu E(\omega )[a_{0}(\omega
)+ia_{1}(\omega )]/2 \\
& - & e^{i\theta }[a_{0}(\omega )-ia_{1}(\omega )]/2\,, \\
a_{5}(\omega ) & = & -\sqrt{2\kappa }\mu E(\omega )[ia_{0}(\omega
)+a_{1}(\omega )]/2 \\
& + & e^{i\theta }[a_{1}(\omega )+ia_{0}(\omega )]/2\,.
\end{array}
\label{W11}
\end{equation}
As an example, we consider below the output spectrum of the mode
$a_{4}(\omega )$, defined by
\begin{equation}
\left\langle a_{4}^{\dag }(-\Omega )a_{4}(\omega )\right\rangle =2\pi
S_{out}(\omega )\delta (\Omega +\omega ),  \label{W12}
\end{equation}
where $S_{out}$ is the spectrum of the output light, which will be defined later.

\begin{figure*}[tbp]
\centering
\begin{minipage}[c]{.75\textwidth}
\centering
\includegraphics[width=12cm]{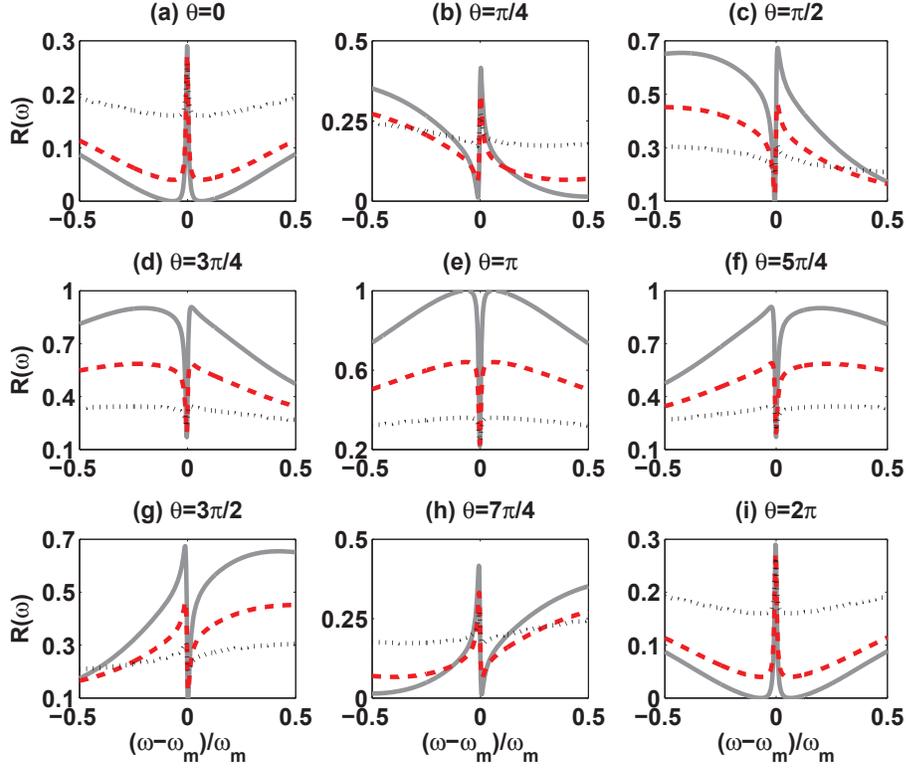}
\end{minipage}
\begin{minipage}[c]{.22\textwidth}
\caption{(Color online) The profiles of the Fano resonance in
variation with $\protect\theta$ and $\mu$, where the parameters take
the same values as in FIG.\ref{symasym}. The curves in each panel: the
gray solid, the red dashed and the black dot correspond to
$\mu=$1.0, 0.6 and 0.2, respectively. } \label{Fano}
\end{minipage}
\end{figure*}

To observe this output spectrum, we introduce the correlation functions
of the quasi-continuum probe field and the input vacuum field,
respectively, \cite {pra-79-013821}
\begin{equation}
\left\langle a_{0}^{\dag}(-\Omega)a_{0}(\omega )\right\rangle =2\pi
S_{in}(\omega)\delta (\omega +\Omega),  \label{W121}
\end{equation}
and
\begin{equation}
\begin{array}{cll}
\left\langle a_{1}(-\Omega)a_{1}^{\dag}(\omega)\right\rangle  & =
& 2\pi \delta (\omega +\Omega).
\end{array}
\label{W13}
\end{equation}
with $S_{in}(\omega)$ being the spectrum of the input field.
Substituting Eq.(\ref{W121}) and Eq.(\ref{W13}) into Eq.(\ref{W12}), the
spectrum of the output field $a_{4}(\omega )$ is given by
\begin{equation}
\begin{array}{cll}
S_{out}(\omega ) & = & R(\omega )S_{in}(\omega ),
\end{array}
\label{W14}
\end{equation}
with
\begin{eqnarray}
R(\omega ) & = & R_{c}(\omega )+\mu ^{2}R_{s}(\omega )+\mu
R_{cs}(\omega )\,,\label{W15}
\\
R_{c}(\omega ) & = & 1/4\,, \\
R_{s}(\omega ) & = & \kappa \mathbf{|}E(\omega
)|^{2}/2\,,\label{W151} \\ R_{cs}(\omega ) & = & \frac{\sqrt{\kappa
}}{2}(\mathbf{\Re }[E(\omega )]\cos \theta +\mathbf{\Im }[E(\omega
)]\sin \theta )\,\label{W152}
\end{eqnarray}
where $R_{s}(\omega)$ indicates the quantum interference in the
output light from the optomechanics (self-interference), while
$R_{cs}(\omega)$ attributes to the compound interference
(cross-interference) between the quasi-continuum light in the lower
arm and the output light from the optomechanics in the upper arm.
Different from the conventional MZI, the optomechanical output light
itself is the result of the interference caused by the OMIT, which
yields $R_{cs}(\omega)$ containing a symmetric component
$\mathbf{\Re}[E(\omega)]$ and an asymmetric one $\mathbf{\Im
}[E(\omega )]$. What is more, the contributions regarding the
symmetric and asymmetric parts can be adjusted by the phase shift at
our will. As a result, the Fano profiles in our scheme are fully
controllable by the phase tuning, whose effect can be observed
directly from the output spectrum of the MZI. As demonstrated in
FIG.\ref{Fano}, the Fano profiles change periodically in an axially
symmetric fashion, where the panels (a-d) are, respectively,
inverted to the panels (e-h).

\begin{figure}[tbp]
\includegraphics[width=7cm]{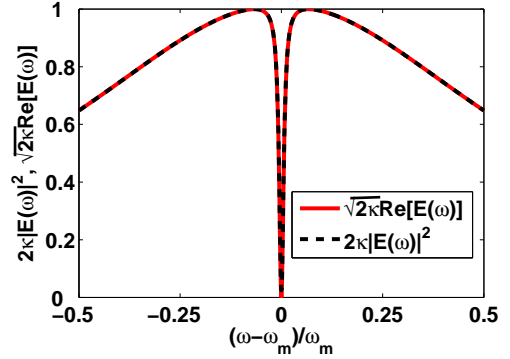}
\caption{(Color online) The comparison between the real part
($\sqrt{2\kappa}Re[E(\omega)]=\Re[\protect\sqrt{2
\protect\kappa} E(\protect\omega)$]) and the square term
($2\kappa|E(\omega)|^2$) of the output light from the optomechanical
cavity. The good agreement implies
$\sqrt{2\kappa}Re[E(\omega)]\simeq 2\kappa|E(\omega)|^2$.}
\label{add}
\end{figure}

To further understand FIG.\ref{Fano}, we plot in FIG.\ref{add} the curves of
$\kappa |E(\omega)|^{2}$ and $\sqrt{\kappa/2}\mathbf{\Re
}[E(\omega)]$, which are in good agreement with each other. This
implies $\kappa |E(\omega )|^{2}\simeq \sqrt{\kappa /2}\mathbf{\Re
}[E(\omega )]$. In fact, we may also understand FIG.\ref{Fano} from
following analysis. In the case of $\cos\theta =-1$, $R(\omega)$
in Eq.(\ref{W15}) can be reduced to
\begin{equation}
\begin{array}{cll}
R(\omega ) & = & 1/4+\mu ^{2}\kappa |E(\omega )|^{2}/2-\mu \sqrt{\kappa /2}
\mathbf{\Re }[E(\omega )] \\
& \simeq & 1/4+(\mu ^{2}-2\mu )\sqrt{\kappa /2}\mathbf{\Re
}[E(\omega )]/2,
\end{array}
\label{W16}
\end{equation}
with $\mu^{2}-2\mu <0$ for $0<\mu\leq 1$. It indicates that the
profile of $R(\omega)$ is very similar to that of the function
$-\sqrt{2\kappa}\mathbf{\Re} [E(\omega)]$ which is inverted with
respect to the function $\sqrt{2\kappa}\mathbf{\Re}[E(\omega)]$.

In conclusion, motivated by recent development of nanotechnology, we
have studied the tunable Fano profile in an optomechanical system by
the phase tuning in the MZI. These phase-controlled Fano
profiles are generated by two kinds of quantum interferences and
vary periodically under our exact control. Since the NAMR has become a
state-of-the-art device for metrology and the Fano profile in our
scheme involves more interferences, we believe that our elaborate
manipulation of the Fano resonance would arouse widespread application,
such as the precision measurement, with the NAMR-based
optomechanics.


We would like to thank Profs. Jin-Hui Wu and Kai-Jun Jiang, and Dr.
Yi Zhang for helpful discussion. The work is supported by the
National Fundamental Research Program of China (Grant Nos.
2012CB922102, 2012CB922104, and 2009CB929604), the National Natural
Science Foundation of China (Grant Nos.
10974225, 60978009,11174027, 11174370, 11274036, 11304366, 11274112, 11204080, 11322542 and 61205108
), and the China Postdoctoral Science Foundation (Grant
Nos. 2013M531771 and 2014T70760).


\begin{thebibliography}{99}
\bibitem{rmp-82-2257} A. E. Miroshnichenko, S. Flach, and Y. S. Kivshar,
Rev. Mod. Phys. \textbf{2010}, 82, 2257.

\bibitem{Cimento-12-154} U. Fano, Nuovo Cimento \textbf{1935}, 12, 154--161.

\bibitem{PR-124-1866} U. Fano, Phys. Rev. \textbf{1961}, 124, 1866--1878.

\bibitem{Nature-440-315} T. Kraemer, M. Mark, P. Waldburger, J. G. Danzl, C.
Chin, B. Engeser, A. D. Lange, K. Pilch, A. Jaakkola, H.-C. Nagerl, and R.
Grimm, Nature (London) \textbf{2006}, 440, 315--318.

\bibitem{prl-107-163604} P. M. Anisimov, J. P. Dowling, and B. C. Sanders
Phys. Rev. Lett. \textbf{2011}, 107, 163604.

\bibitem{pre-71-036626} A. E. Miroshnichenko, S. F. Mingaleev, S. Flach, and
Y. S. Kivshar, Phys. Rev. E \textbf{2005}, 71, 036626.

\bibitem{pre-74-046603} S. F. Mingaleev, A. E. Miroshnichenko, Y. S.
Kivshar, and K. Busch, Phys. Rev. E \textbf{2006}, 74, 046603.

\bibitem{NL-9-1663} N. Verellen, Y. Sonnefraud, H. Sobhani, F. Hao, V. V.
Moshchalkov, P. V. Dorpe, P. Nordlander, and S. A. Maier, Nano Lett. \textbf{%
2009}, 9, 1663--1667.


\bibitem{pra-79-013809} A. E. Miroshnichenko, Y. Kivshar, C. Etrich, T.
Pertsch, R. Iliew, and F. Lederer, Phys. Rev. A \textbf{2009}, 79, 013809.

\bibitem{prl-90-084101} S. Flach, A. E. Miroshnichenko, V. Fleurov, and M.
V. Fistul, Phys. Rev. Lett. \textbf{2003}, 90, 084101.

\bibitem{PhysScr-74-259} Y. S. Joe, A. M. Satanin, and C. S. Kim, Phys. Scr.
\textbf{2006}, 74 259--266.

\bibitem{rmp-79-1217} R. Hanson, L. P. Kouwenhoven, J. R. Petta, S. Tarucha,
and L. M. K. Vandersypen, Rev. Mod. Phys. \textbf{2007}, 79 1217--1265.



\bibitem{arxiv} K. Qu, and G. S. Agarwal, Phys. Rev. A \textbf{2013}, 87, 063813.

\bibitem{oe-21-6601}Q. Zhang, J. J. Xiao, X. M. Zhang, Y. Yao, and H. Liu, Opt. Exp. \textbf{2013} 21 6601-6608.

\bibitem{ajp-95-2682} K. L. Ekinci, Y. T. Yang, and M. L. Roukes, J. Appl.
Phys. \textbf{2004}, 95, 2682.

\bibitem{pra-86-053806} J. -Q. Zhang, Y. Li, M. Feng, and Y. Xu, Phys. Rev.
A \textbf{2012}, 86, 053806.

\bibitem{prl-95-221101} B. Abbott \textit{et al.} Phys. Rev. Lett. \textbf{2005
}, 95, 221101.

\bibitem{Physics-2-40} F. Marquardt, and S. M. Girvin, Physics \textbf{2009}, 2,
40.

\bibitem{prl-108-120801} S. Forstner, S. Prams, J. Knittel, E. D. van
Ooijen, J. D. Swaim1, G. I. Harris, A. Szorkovszky, W. P. Bowen, and H.
Rubinsztein-Dunlop, Phys. Rev. Lett. \textbf{2012}, 108, 120801.

\bibitem{prl-88-120401} S. Mancini, V. Giovannetti, D. Vitali, and P.
Tombesi, Phys. Rev. Lett. \textbf{2002}, 88, 120401.

\bibitem{prl-98-030405} D. Vitali, S. Gigan, A. Ferreira, H. R. Bohm, P.
Tombesi, A. Guerreiro, V. Vedral, A. Zeilinger, and M. Aspelmeyer, Phys.
Rev. Lett. \textbf{2007}, 98, 030405.

\bibitem{prl-101-200503} M. J. Hartmann and M. B. Plenio, Phys. Rev. Lett.
\textbf{2008}, 101, 200503.

\bibitem{prl-107-063601} P. Rabl, Phys. Rev. Lett. \textbf{2011}, 107, 063601.

\bibitem{prl-107-063602} A. Nunnenkamp, K. Borkje, and S. M. Girvin, Phys.
Rev. Lett. \textbf{2011}, 107, 063602.

\bibitem{OL-38-353} H. Xiong, L.-G. Si, X.-Y. Lv, X. Yang and Y. Wu, Opt. Lett. \textbf{2013}, 38, 353-355.

\bibitem{pra-81-041803} G. S. Agarwal, and S. Huang, Phys. Rev. A \textbf{2010}, 81, 041803.

\bibitem{Science-330-1520} S. Weis, R. Riviere, S. Deleglise, E. Gavartin,
O. Arcizet, A. Schliesser, and T. J. Kippenberg, Science \textbf{2010}, 330, 1520.

\bibitem{pra-86-013815} H. Xiong, L.-G. Si, A.-S. Zheng, X. Yang and Y. Wu, Phys. Rev. A \textbf{2012}, 86, 013815.

\bibitem{Nature-472-69} A. H. Safavi-Naeini, T. P. M. Alegre, J. Chan, M.
Eichenfield, M. Winger, Q. Lin, J. T. Hill, D. Chang, and O. Painter, Nature
(London) \textbf{2011}, 472, 69--73.

\bibitem{pra-85-021801} G. S. Agarwal and S. Huang, Phys. Rev. A \textbf{2012}, 85, 021801.

\bibitem{JPConnerade} J. P. Connerade, \emph{Highly Excited Atoms}; Cambridge University Press: Cambridge, 1998.

\bibitem{pra-84-033828} Y. Xu, and A. E. Miroshnichenko, Phys. Rev. A
\textbf{2011}, 84, 033828.

\bibitem{D.F.Walls-Quantum Optics} D. F. Walls and G. J. Milburn, Quantum
Optics; Springer-Verlag: Berlin, 1994.

\bibitem{Nature-460-724} S. Groblacher, K. Hammerer, M. Vanner, and M.
Aspelmeyer, Nature (London) \textbf{2009}, 460, 724.

\bibitem{Nature-452-72} J. D. Thompson, B. M. Zwickl, A. M. Jayich, F.
Marquardt, S. M. Girvin, and J. G. E. Harris, Nature (London) \textbf{2008}, 452,
72.

\bibitem{quantumoptics} C. C. Gerry, and P. Knight, Introductory Quantum
Optics; Cambridge University Press: New York, 2005

\bibitem{pra-79-013821} S. Huang, and G. S. Agarwal, Phys. Rev. A \textbf{2009}%
, 79, 013821.

\bibitem{jpb-34-1953} Ph. Durand, I. Paidarova, and F. X. Gadea, J. Phys. B:
At. Mol. Opt. Phys. \textbf{2001}, 34, 1953.
\end{thebibliography}
\end{document}